\documentclass[aps,twocolumn,prl]{revtex4}
\usepackage{epsfig}
\usepackage{graphicx}
\usepackage{amsmath}

\begin{document}


\title{Electron-hole asymmetry in magnetic properties of lightly doped 
high-$T_{\rm C}$ superconductors: a quantum Monte Carlo study}

\author{Seiji Yunoki}
\email{yunoki@sissa.it}
\author{Sandro Sorella}
\email{sorella@sissa.it}

\affiliation{International School for Advanced Studies (SISSA), 
via Beirut 4, 34014 Trieste, Italy}



\begin{abstract}

Using a recently developed variational quantum Monte Carlo method, 
magnetic properties of high-$T_{\rm C}$ superconductors are studied 
at zero temperature ($T$), by numerical simulations on the 2D $t$-$J$ model.   
Our focus here is to explore the difference in the properties of $p$-type 
and $n$-type cuprates as a function of the carrier concentrations close 
to half filling. As observed experimentally, it is found that the 
antiferromagnetically ordered phase persists even for a small, yet finite 
amount of carrier doping, and that this phase is more robust in the 
electron doped case.

\end{abstract}

%
%
\maketitle


There have been stimulating studies on high-$T_{\rm C}$ superconductors 
to explore the underlying mechanism and unusual normal state properties. 
Experimentally much attention has been recently devoted  
to rather lightly doped $p$- and $n$-type cuprates~\cite{exp}. 
This is partially because high quality samples have become now available, and 
because it is now clear that understanding the evolution of the electronic 
state with 
carrier doping from undoped Mott insulator is crucial to 
clarify the nature of the systems. Theoretically it is still in the middle of 
extensive debate. It is widely accepted however that a 2D $t$-$J$-like 
model can describe low-energy 
physics of cuprates defined by the following Hamiltonian~\cite{rice}; 
\begin{eqnarray}
H &=& J \sum_{ \langle i,j \rangle } \left( {\bf S}_i \cdot {\bf S}_j 
   -n_in_j/4 \right)
   - t \sum_{ \langle i,j \rangle \sigma } \left( {\tilde c}^{\dag}_{i,\sigma} 
              {\tilde c}_{j,\sigma} + {\rm H.c.}  \right ) \nonumber \\
  &&- t' \sum_{ \langle\langle i,j \rangle\rangle \sigma } 
  \left( {\tilde c}^{\dag}_{i,\sigma} 
              {\tilde c}_{j,\sigma} + {\rm H.c.}  \right ).
\label{tj}
\end{eqnarray}
Here the notation is conventional (see {\it e.g.}, Ref~\cite{sandro}). 
Introduction of next nearest-neighbor hopping $t'$ deserves some 
explanation; small cluster calculations have shown that the 
$t$-$J$ model indeed capture low energy physics of more involved 
multi-band models with $t'/t$ negative (positive) for hole-doped 
(electron-doped) case, at least for small carrier 
densities $x$~\cite{eskes}. We remind that this sign difference 
originates from the fact that the parent cuprates are charge transfer 
insulators, and therefore a ``carrier'' represents in the model 
a Zhang-Rice singlet (an inert Cu $d^{10}$ configuration) for 
the hole (electron) doped system. 
It is also now widely accepted that, in the half-filling undoped case, 
the model shows antiferromagnetic (AFM) long-range order 
at $T=0$.  It is very interesting therefore to study theoretically 
how strong this 
AFM phase is against carrier doping and how the robustness of the phase 
depends on the nature of doped carriers.

There exist several numerical studies on small clusters ($\le$ 26 sites) 
indicating that at finite $x$ close to half-filling positive (negative) 
$t'/t$ 
enhances (degrades) AFM correlations compared to $t'=0$~\cite{good}. 
Although these results might give the correct trend at short distances, 
it is difficult within such small cluster calculations to really assess the 
existence of long-range order. For this purpose we have studied this 
issue using quantum Monte Carlo techniques, which allows much larger 
sizes ($\simeq 100$ sites).


\begin{figure}[hbt]
\begin{center}\leavevmode
\includegraphics[width=0.8\linewidth]{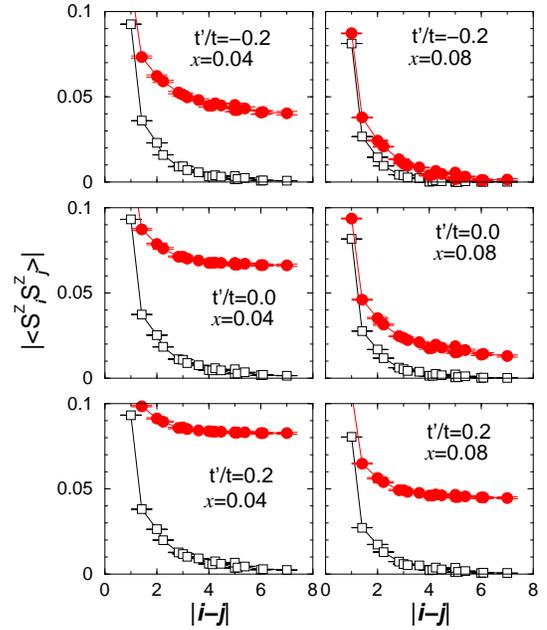}
\caption{Absolute value of real-space spin correlation functions for 2D 
$t$-$t'$-$J$ model with $J/t=0.3$ and $t'/t=-0.2$ (top), 
$0.0$ (middle), and $0.2$ (bottom). Carrier densities $x$ are 0.04 (left) 
and $0.08$ (right) on a tilted square lattice of $\sqrt{98}\times\sqrt{98}$ . 
Open squares and solid circles are for VMC and FN, respectively. 
}
\label{sz1}
\end{center}
\end{figure}

To study ground sate properties of the 2D $t$-$t'$-$J$ model (\ref{tj}), 
we employ a recently developed variational quantum Monte Carlo method by 
Sorella~\cite{sandro,ceperley}. Results presented here are computed by 
variational and fixed node methods (VMC and FN, respectively), and 
the details on how to minimize the energy for a variational wave function 
with many parameters are found in Ref.~\cite{sandro}. 
We just mention here that we employ the most general singlet variational 
wave function of the Jastrow-BCS form, with a pairing function of $d$-wave 
symmetry. The Jastrow factor, which includes the Gutzwiller projector, 
makes the wave function of the RVB type. 
After all the possible variational parameters are optimized, we obtain a 
fairly good guiding function that is used for FN. 
Also note that the variational energy of FN is lower than the one of 
VMC~\cite{ceperley}.

\begin{figure}[hbt]
\begin{center}\leavevmode
\includegraphics[width=0.8\linewidth]{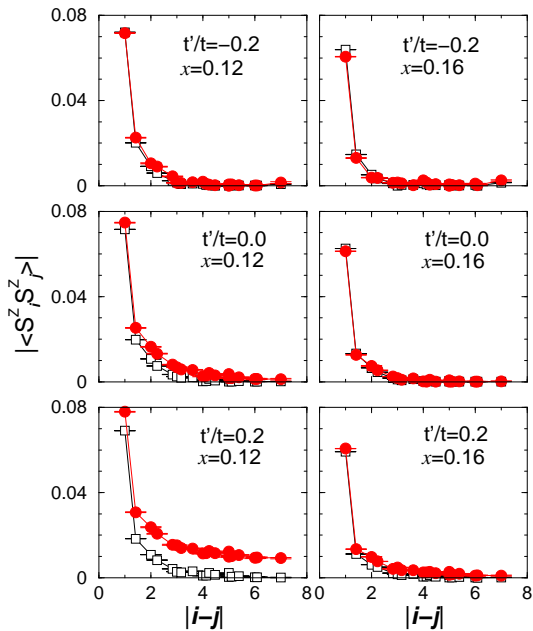}
\caption{Same as in Fig.~\ref{sz1}, but $x=0.12$ (left) and $0.16$ (right).}
\label{sz2}
\end{center}
\end{figure}

Main results are summarized in Figs.~\ref{sz1} and \ref{sz2} where 
real-space spin-spin correlation functions are given for various $t'/t$ 
and doped carrier concentrations $x$~\cite{note} at $J/t=0.3$. 
These results provide clear evidence that strong spin correlations 
remain finite at the largest distances for finite carrier concentrations $x$. 
The critical $x_{\rm c}$ above which long range order disappears 
apparently depends on the parameters used; for the 
present set of $t'/t=-0.2$, $0$, and $0.2$, $x_{\rm c}\simeq 0.06$, $0.08$, 
and $0.12$, respectively. Of particular interest is the result that 
the spin correlations are enhanced (degraded) when positive (negative) 
$t'/t$ is incorporated into the model, corresponding to electron (hole) 
doing. Our results are consistent with the previous 
studies~\cite{good,elbio} and experimental observations~\cite{mag}.


In conclusion, magnetic properties of the lightly doped 2D $t$-$J$ 
model with next nearest neighbor hopping $t'$ are numerically studied 
to understand the different behaviors as a function of carrier 
concentrations for hole ($t'/t<0$) and electron ($t'/t>0$) doping. 
Our large cluster calculations clearly show that AFM long-range order 
exists away from half-filling and is more robust for the electron doped 
case. These results compare very well with experimental observations. 
More detailed analyses and other properties will be presented in a 
separate paper~\cite{yunoki}. 

This work was supported by MIUR-COFIN 2001.
  
%
%
%

%
%


\begin{thebibliography}{9}
\bibitem{note} Note that $x=0$ corresponds to undoped systems.
\end{thebibliography}
\end{document}